# Magnetic properties of Sn-substituted Ni-Zn ferrite: synthesized from nano-sized powders of NiO, ZnO, $Fe_2O_3$ and $SnO_2$

M.A. Ali[1], M.M. Uddin[1,*], M.N.I. Khan[2], F.-U.-Z. Chowdhury[1], S.M. Hoque[2], S.I. Liba[2]

[1]Department of Physics, Chittagong University of Engineering and Technology (CUET), Chittagong 4349, Bangladesh

[2]Materials Science Division, Atomic Energy Center, Dhaka 1000, Bangladesh

**Abstract:**
A series of $Ni_{0.6-x/2}Zn_{0.4-x/2}Sn_xFe_2O_4$ ($x$ = 0.0, 0.05, 0.1, 0.15, 0.2 and 0.3) (NZSFO) ferrite composites have been synthesized from nano powders using standard solid state reaction technique. The spinel cubic structure of the investigated samples has been observed by the X-ray diffraction (XRD). The magnetic properties such as saturation magnetization ($M_s$), remanent magnetization ($M_r$), coercive field ($H_c$) and Bohr magneton ($\mu_B$) are calculated from the hysteresis loops. The value of $M_s$ is found to decrease with increasing Sn content in the samples. This change has been successfully explained by the variation of $A$-$B$ interaction strength due to Sn substitution in different sites. The compositional stability and quality of the prepared ferrite composites have also been endorsed by the fairly constant initial permeability ($\mu^/$) over a wide range of frequency region. The decreasing trend of $\mu^/$ with increasing Sn content has been observed. Curie temperature ($T_C$) has found to increase with the increase in Sn content. Wide spread frequency utility zone indicates that the NZSFO can be considered as a good candidate for use in broadband pulse transformer and wide band read-write heads for video recording. The abnormal behavior for $x$ = 0.05 has been explained with existing theory.

**Keywords:** Magnetic properties, Saturation magnetization, Permeability, Curie temperature.

PACS: 75; 07.55.-w; 75.50.Gg; 75.60.Ej; 75.50.Ss

## 1. Introduction

Over the last few decades, scientific community has paid significant attention to the spinel ferrites due to their fascinating properties to meet the requirements in various applications. No other magnetic materials can replace the ferrites due to their low price, availability, stability and have an extensive use of technological application in transformer, high quality filters, high and very high frequency circuits and operating devices. The Ni-Zn ferrites became an important candidate to use it in the high frequency applications due to their high electrical resistivity, high permeability, compositional stability and low eddy current losses [1-6]. The uniqueness of Ni-Zn ferrites is motivating numerous researchers to look forward that they can open the way for commercial applications of these materials and new types of ferrites are unveiling with excellent properties for practical application. The properties of Ni-Zn ferrites can be tailored by altering chemical composition, preparation methods, sintering temperature ($T_s$) and impurity element or levels and the reports regarding these issues are available in the literatures [5-28]. Recently, many researchers reported the structural, magnetic and electrical properties of Ni-Zn ferrite and/or substituted Ni-Zn ferrites in bulk [6, 12, 13, 19-21, 26-28] and nano forms [16, 22-24].

The properties of Ni-Zn ferrites can be changed remarkably by substitution of tetravalent ions such as $Ti^{4+}$, $Sn^{4+}$. Investigations on the substitution of $Sn^{4+}$ have been reported by many researchers [6, 9-11]. We have reported the structural, morphological and electrical properties of Sn-substituted Ni-Zn ferrites [6]. Das *et al* reported the variation of lattice parameter, saturation magnetization and Curie temperature with $Ti^{4+}$, $Zr^{4+}$ substitution, and $Sn^{4+}$ in Ni-Zn ferrite compositions synthesized by chemical method [9]. The $Sn^{4+}$ substituted $Ni_{1-y}Zn_yFe_2O_4$ (*y* = 0.3, 0.4) ferrite samples were prepared in an oxidizing atmosphere using the solution technique and studies on Mössbauerand magnetization properties have been investigated by Khan *et al* [10]. The magnetic hysteresis and the thermal variation of magnetic parameters in $Sn^{4+}$ doped Ni-Zn ferrites prepared by standard ceramic technique have also been reported by

Maskar *et al* [11]. We have synthesized the $Sn^{4+}$ substituted Ni-Zn ferrites using nano powders (which is different from Maskar *et al* work [11]) by standard ceramic technique. Another significant dissimilarity is that they have doped Sn in the Ni-Zn ferrites, where as we have simultaneously substituted Sn for both Ni and Zn. The characterization and the frequency dependence of magnetic properties of Ni-Sn-Zn ferrites provide the way to classify the ferrites for particular applications. The information would be very noteworthy for the scientific community in this regard.

To the best of our knowledge, such type of study of Ni-Zn ferrites prepared from nano powders has not been reported yet. Here, we report the magnetic properties of Sn-substituted $Ni_{0.6}Zn_{0.4}Fe_2O_4$ ferrites prepared from nano-sized raw materials by the solid state reaction technique.

## 2. Materials and methods

Solid state reaction route was followed to synthesize Sn-substituted Ni-Zn ferrite, $Ni_{0.6-x/2}Zn_{0.4-x/2}Sn_xFe_2O_4$ ($0.0 \leq x \leq 0.30$) (NZSFO). High purity (99.5%) (US Research Nanomaterials, Inc.) oxides of nano powders were used as raw materials. The particle size of nickel oxide (NiO), zinc oxide (ZnO), iron oxide ($Fe_2O_3$) and tin oxide ($SnO_2$) are 20-40, 15-35, 35-45 and 35-55 nm, respectively. The detail of preparation technique has been described elsewhere [5, 6]. The phase formation and surface morphology of the synthesized samples were carried out by the X-ray diffraction (XRD) using Philips X'pert PRO X-ray diffractometer (PW3040) with $CuK_\alpha$ radiation ($\lambda$ = 1.5405 Å) and scanning electron microscope (SEM), respectively. The magnetic properties (*M-H* curve, saturation magnetization, $M_s$; coercive field, $H_c$; and Bohr magneton; $\mu_B$) have been elucidated by the vibrating sample magnetometer (VSM) (Micro Sense EV9) with a maximum applied field of 10 kOe. Frequency and temperature dependent permeability were investigated by using

Wayne Kerr precision impedance analyzer (6500B). An applied voltage of 0.5 V with a low inductive coil was used to measure permeability.

## 3. Results and discussion

*3.1. XRD analysis*

The X-ray diffraction (XRD) pattern of Sn-substituted Ni-Zn ferrites with the chemical composition of $Ni_{0.6-x/2}Zn_{0.4-x/2}Sn_xFe_2O_4$ (NZSFO) are shown in Fig. 1. The XRD spectra were indexed and fcc cubic phase was identified. The structural parameters are calculated from the XRD data and have been discussed in Ali *et al* [6]. The lattice constants are calculated from the XRD data and represented in Table 1.

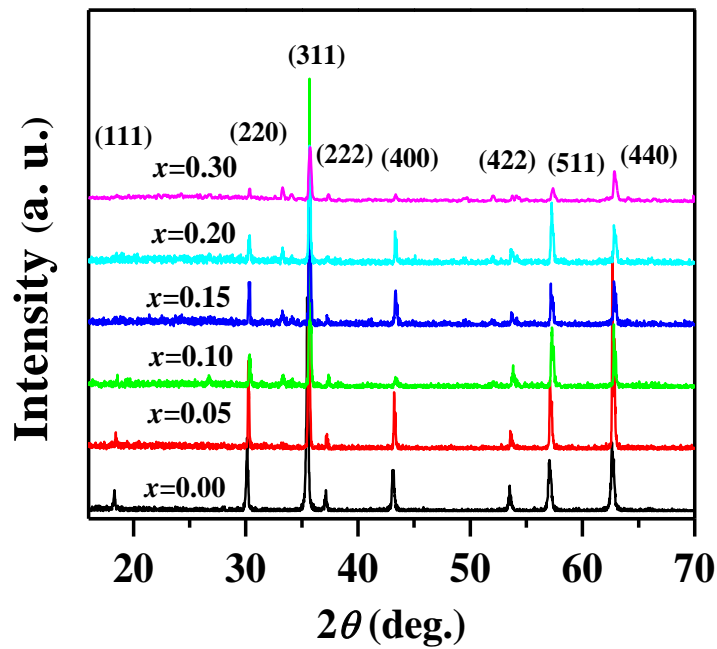

**Fig.1:** The X-ray diffraction pattern of the NZSFO ($x$ = 0.0, 0.05, 0.1, 0.15, 0.2, 0.3 and 0.4) ferrites samples [6].

The distances between the magnetic ions at tetrahedral (*A*) and octahedral (*B*) sites have been calculated using the equation: $L_A = a\frac{\sqrt{3}}{4}$ and $L_B = a\frac{\sqrt{3}}{2}$. The values are also depicted in Table 1. The hopping lengths of $L_A$ and $L_B$ decrease with increasing Sn concentration might because of lattice parameters of the Ni-Zn ferrites decrease with increasing $Sn^{4+}$ concentration.

## 3.2 Magnetic properties

The plots of applied magnetic field $H$ (up to 10 kOe) dependent magnetization at room temperature, of $Ni_{0.6-x/2}Zn_{0.4-x/2}Sn_xFe_2O_4$ ($x$ = 0.0, 0.05, 0.1, 0.15, 0.2, and 0.3) ceramics sintered at 1300 °C, are shown in Fig. 2.

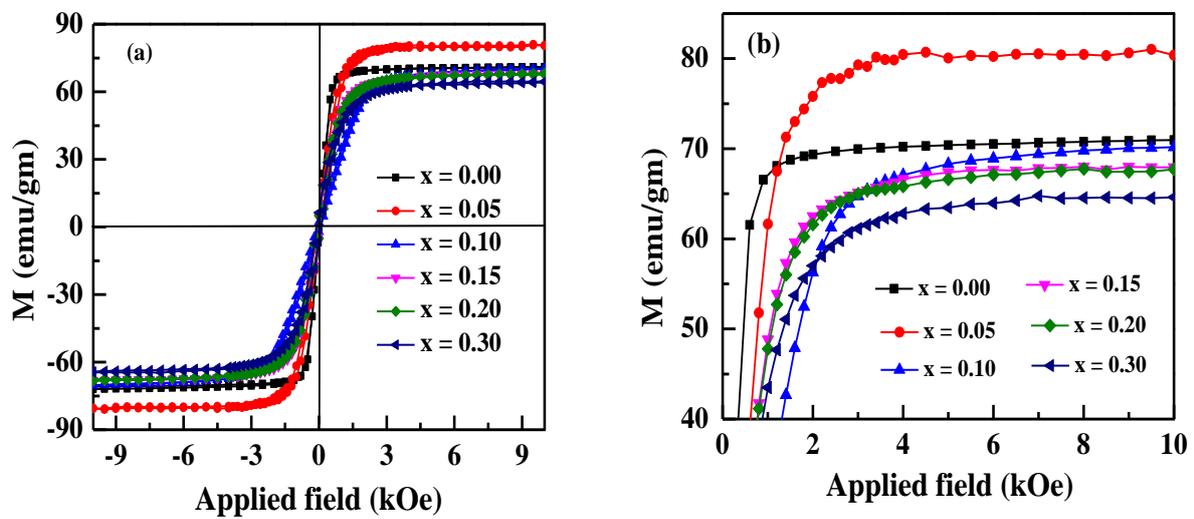

**Fig. 2:** (a) *M–H* loops of the NZSFO ferrite samples, (b) magnification of upper saturated part of the *M-H* loops.

The value of magnetization increases with increasing applied magnetic field up to a certain field above which the sample becomes saturated. The saturation magnetization ($M_s$), coercive field ($H_c$), remanent magnetization, $M_r$, and Bohr magneton, $\mu_B$, are also calculated from the measured magnetic hysteresis loop and are presented in Table 1. It is seen that the values of $H_c$ of the Sn-substituted samples (NZSFO) are larger than that of the parent (NZFO) and it could be inferred that the prepared ferrites are not reasonably soft in nature.

The variation of saturation magnetization ($M_s$) with Sn content of the NZSFO is shown in Table 1. The value of $M_s$ shows decreasing trend with increasing $x$ except $x = 0.05$ where the $M_s$ increases slightly. This characteristic can be understood in terms of exchange interactions of cations in the samples. However, the *AB* interactions are generally dominant in the ferrites, *AA* and *BB* interactions can no longer be ignored. The magnetic moment of the Ni ion is affected by the modified strength of $Ni^{2+} \leftrightarrow O^{2-} \leftrightarrow Ni^{2+}$ interactions due to the presence of non-magnetic ions Sn and Zn in the samples.

**Table 1:** The lattice constants ($a_{expt}$), average grain size ($D_g$), magnetic cation hopping length ($L_A$ and $L_B$), saturation magnetization ($M_s$), coercive field ($H_c$), remanent magnetization ($M_r$) and Bohr magneton ($n_B$) of NZSFO for different $x$.

| Composition $x$ | $a_{expt}$ (Å) | $D_g$ (μm) | $L_A$ (Å) | $L_B$ (Å) | $M_s$ (emu/gm) | $H_c$ (Oe) | $M_r$ (emu/gm) | $\mu_B$ |
|---|---|---|---|---|---|---|---|---|
| 0.00 | 8.39311 | 07.8 | 3.6343 | 7.2686 | 72.73<br>60.0 [9]<br>96.0 [11] | 1.26 | 0.16 | 3.09 |
| 0.05 | 8.38996 | 10.1 | 3.6329 | 7.2659 | 80.82 | 67.25 | 5.00 | 3.48 |
| 0.10 | 8.37546 | 18.8 | 3.6266 | 7.2533 | 70.23 | 62.41 | 2.07 | 3.07 |
| 0.15 | 8.38137 | 21.0 | 3.6292 | 7.2584 | 68.09 | 71.82 | 4.55 | 3.02 |
| 0.20 | 8.37665 | 30.1 | 3.6271 | 7.2543 | 68.07 | 78.10 | 5.16 | 3.06 |
| 0.30 | 8.34531 | 34.8 | 3.6136 | 7.2272 | 64.47 | 94.47 | 6.40 | 2.99 |

It is assumed that the Sn ions occupy tetrahedral (*A*) sites initially at lower Sn concentrations however, it resides in *B*-sites at higher Sn concentrations leading the reduction of *A-A* interactions. Consequently, the value of net magnetic moment, $\overline{M} = \overline{M}_B - \overline{M}_A$, increases in the samples. On further increase of substituting Sn ions, they enter into *B*-sites and pushing some $Fe^{3+}$ ions into *A*-sites resulting the magnetic ion density in the *B* sub-lattice decreases. The concentration of $Fe^{3+}$ ions decreases in the *B* sub-lattice while it increases for *A* sub-lattice thereby the net magnetic moment of the ferrite diminishes. Our calculated values of $M_s$ shown in Table 1 are compared with the reported values [9, 11]. Das *et al* [9] have reported the value of $M_s \sim 60$ emu/gm for $x = 0.0$ and observes the $M_s$ decreases with increasing Sn concentration up to 5 wt% in the Ni-Zn ferrites. Maskar *et al* [11] have also observed that the value of $M_s \sim 96$ emu/gm for $x=0.0$ and noticed the lower $M_s$ value for further substitution in Ni-Zn ferrites. We have found the value of $M_s \sim 72.2$ emu/gm, and a decreasing trend with increasing Sn content up to 40 wt% is also observed, except for $x = 0.05$. As mentioned in Table 1, the $M_s$ for $x = 0.0$ is found to differ from the reported values of other researchers (ref. [9] and [11]). This discrepancy in $M_s$ value might be due to dissimilarity in sample preparation techniques and conditions employed.

The Sn content dependence of the coercive field of NZSFO is depicted in Table 1. It shows that the $H_c$ value increases with increasing Sn content that can be elucidated by the Brown's relation: $H_c = 2K_1/\mu_0 M_s$, where $K_1$ is the anisotropy constant and $\mu_0$ is the permeability of free space. As per relation, the value of $H_c$ is found to increase with the decrease in the value of $M_s$. Furthermore, Stoner–Wohlfarth single-domain theory proved that the $H_c$ increases with the increase of grain size [29]. The grain sizes of the prepared samples (NZFO and NZSFO) are also found to increase with Sn contents [6]. Therefore, it is expected to increase the value of $H_c$ with the increase of $x$ in the prepared samples. It could be noted that the values of $H_c$ for the Sn-substituted samples are comparatively large. The large value of $H_c$ for the Sn-substituted samples can be explained by the following equations. $H_c = \frac{\pi r}{M_s}\left(K_1/A\right)^{1/2}$, where *A* is the exchange stiffness constant, $K_1$ is the anisotropy constant and *r* is the radius of the spherical pores [30]. In general, $H_c$ varies directly with porosity and anisotropy; and inversely with grain size [30]. Thus, the $H_c$ appears to be influenced by saturation magnetization and $K_1$, in addition to the microstructure.

The porosity of NZSFO increases almost linearly with Sn concentration (~27-34%) while the porosity for NZFO is around 19% [6]. The grain size of the prepared samples (Table 1) is found to increase with Sn contents from 10 to 34 μm for (0.05 ≤ x ≤ 0.3) while the grain size of NZFO is only 7.8 μm. In addition, the increase in $T_c$, suggests that the value of $K_1$ is also increased with Sn contents. It can be recalled that the value of $M_s$ decreases with Sn content. As a result, the value of $H_c$ is much larger for Sn-substituted samples (NZSFO). The values of $M_r$ and $\mu_B$ as a function of Sn concentration are also presented in Table 1. The mechanism for the variation of $\mu_B$ and $M_r$ is closely related to the $M_s$ and $H_c$, respectively.

Fig. 3 represents the real part of initial permeability ($\mu'$) and imaginary part of the initial permeability ($\mu''$) over the frequency range from 1 kHz to 100 MHz for the NZSFO for different Sn concentration. The $\mu'$ and $\mu''$ of the $\mu^*$ have been calculated using the following relations: $\mu' = L_s/L_0$ and $\mu'' = \mu'.\tan\delta$, where $L_s$ is the self-inductance of the sample core and $L_0 = \frac{\mu_0 N^2 S}{\pi \bar{d}}$ can be derived geometrically, where $L_0$ is the inductance of the winding coil without the sample core, $N$ is the number of turns of the coil ($N = 5$), $S$ is the area of cross-section of the toroidal sample as given below: $S = d \times h$ and $d = \frac{d_2 - d_1}{2}$, here $d_1$ = inner diameter, $d_2$ = outer diameter and $h$=heightand also $\bar{d}$ is the mean diameter of the toroidal sample ($\bar{d} = \frac{d_2 + d_1}{2}$). The real part of permeability $\mu'$ decreases with the frequency and the imaginary part of permeability $\mu''$ exhibits a peak, which is related to the relaxation phenomena. It is seen that the $\mu'$ remains almost constant until the frequency is raised to a certain value and then drops to very low values at higher frequencies.The fairly constant $\mu'$ values with a wide range frequency region is known as the zone of utility of the ferrite that demonstrate the compositional stability and quality of ferrites prepared by conventional double sintering route. This characteristics is anticipated for various applications such as broadband pulse transformer and wide band read-write heads for video recording [31]. The value of $\mu''$ gradually increases with the frequency and become maximum at a certain frequency, where $\mu'$ rapidly decreases. This feature is well known as the ferromagnetic resonance [32]. At higher doping concentration, the permeability value is lower and the frequency of the onset of ferromagnetic resonance is higher that is in good agreement with Snoek's limit $f_r \mu' =$ constant, where $f_r$ is the resonance frequency of domain wall motion, above which $\mu'$ decreases [33].

Variation of $\mu'$ with Sn concentration at 1 MHz frequency is shown in Fig. 4(a). The decrease in the initial permeability of the Ni–Zn ferrites can be explained using the following equation $\mu' = \frac{M_s^2 D}{\sqrt{K_1}}$, where $\mu'$ is the initial permeability, $M_s$ the saturation magnetization, D the average grain size and $K_1$ the magneto-crystalline anisotropy constant. As $\mu'$ is proportional to the square of the saturation magnetization and saturation magnetization is decreased with the increase in Sn concentration, the value of $\mu'$ is expected to be decreased. Tetravalent $Sn^{4+}$ ions have a strong octahedral-site preference, and the saturation magnetization decreased with the increasing $Sn^{4+}$ substitution due to the weaker A–O–B super-exchange interaction results the value of $\mu'$ decreases [34]. Fig. 4(b) shows the relative quality factor NZSFO. The peak corresponding to maxima in Q-factor shifts to a higher frequency range as Sn content increases. Q-factor has the maximum value of $5.2 \times 10^3$ at $f = 20$ MHz for the $x = 0.05$ sample. The Q-value depends on the ferrite microstructure, e.g. pore, grain size, etc.

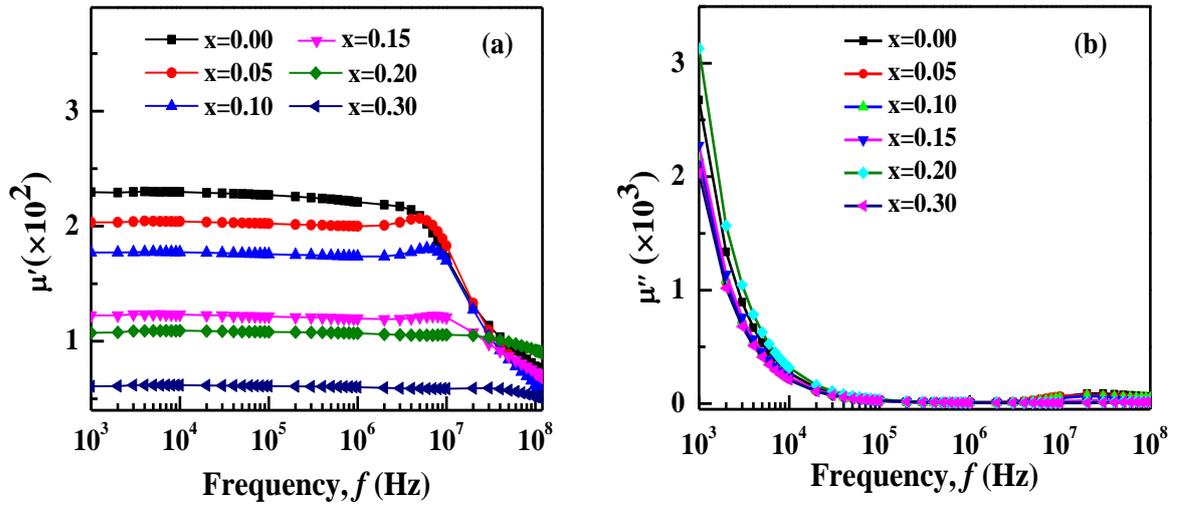

**Fig. 3:** The frequency dependence of permeability (a) real part (b) imaginary part of the NZSFO for different Sn concentration.

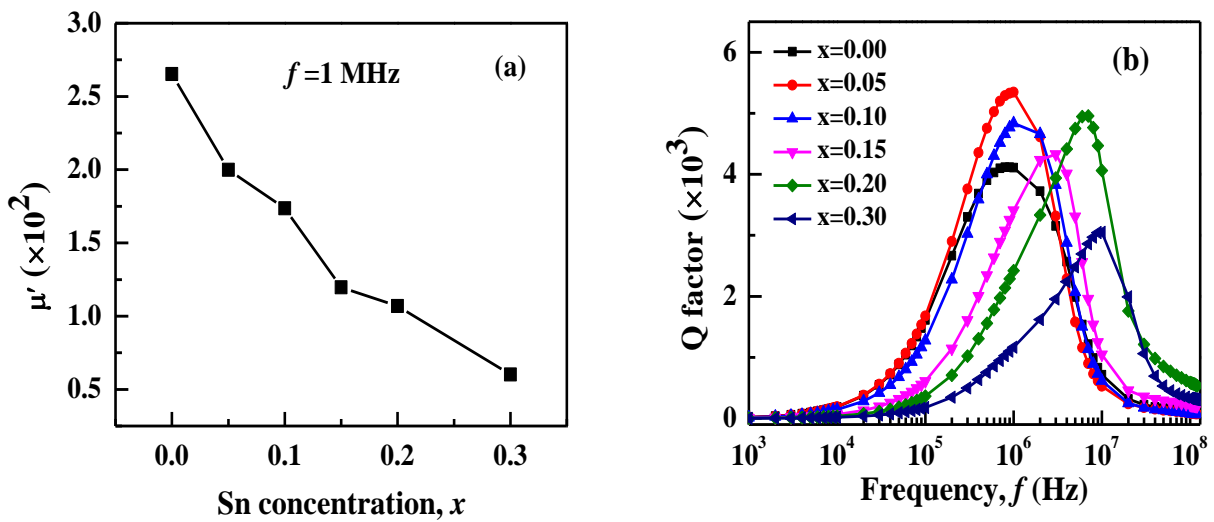

**Fig. 4:** Variation of (a) $\mu'$ with Sn concentration at 1 MHz frequency, (b) variation of Q-factor with frequency for different Sn concentration in the NZSFO ferrites.

Curie temperature ($T_c$) is the transition temperature above which the ferrite material loses its magnetic properties. Temperature dependence of initial permeability, $\mu'$ of the toroid shaped sample of NZSFO at constant frequency 1 MHz of an AC signal is shown in Fig. 5 (a). The initial permeability increases rapidly with increasing temperature and then drops off sharply near the transition temperature known as $T_c$ showing the Hopkinson effect [35]. A significant peak is obtained near the $T_c$ where the value of $K_1$ becomes almost negligible. At the $T_c$, complete spin disorder takes place, i.e., a ferromagnetic material converts to a paramagnetic material. The $T_c$ gradually increases with increasing Sn concentration excluding $x = 0.05$ where it is decreased moderately as shown in Fig. 5 (b).

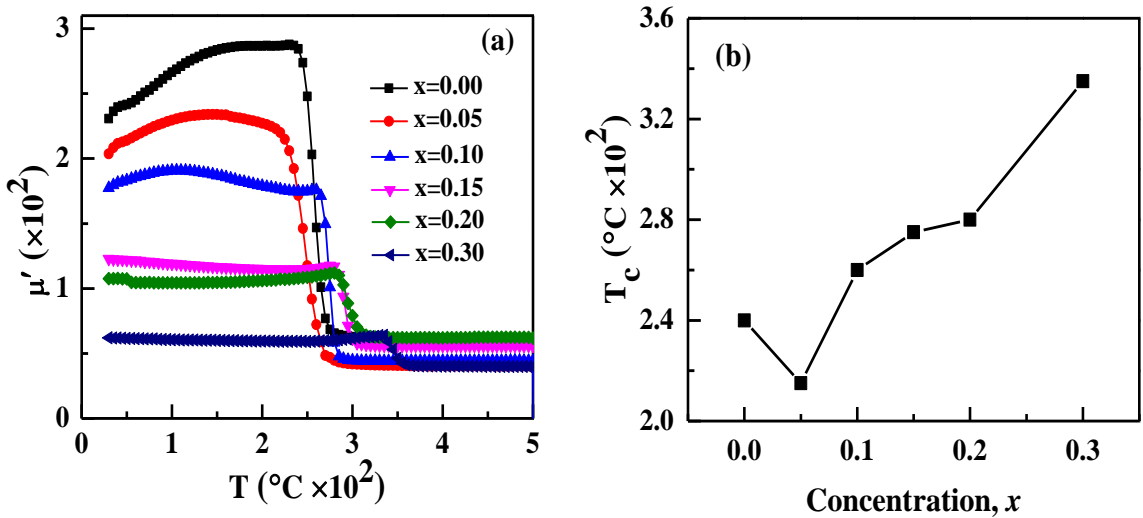

**Fig. 5**: (a) Temperature dependent initial permeability ($\mu'$) for different Sn concentration, (b) variation of the Curie temperature ($T_c$) as a function of Sn concentration in the NZSFO ferrites.

It can be explained as follows; initially the dopant cations are assumed to occupy tetrahedral (*A*) sites and it is entered into *B* sites due to further increasing of dopant cations thereby pushing some $Fe^{3+}$ ions to *A*-sites resulting the magnetic ion density decreases in the *B* sub-lattice [9]. The increase in the magnetic ions in the *A* sites increases the *A-B* interaction, consequently increasing the $T_c$ (Fig. 5 (b)). However, the mechanism of the $T_c$ decreasing in particular point has not been understood yet.

Finally, from the Table 1, it is evident that the sample with $x = 0.05$ shows unusual results. Moreover, the $T_c$ values are fairly linear with Sn content except for $x = 0.05$. This unusual behavior might be explained assuming that initially at lower Sn concentrations ($0 \leq x \leq 0.1$) Sn ions occupy tetrahedral (*A*) sites whilst these ions reside in octahedral (*B*) sites at higher Sn concentrations [9]. However, this can be confirmed by other investigations like, neutron diffraction, but unfortunately we are unable to perform such investigation and left this issue to other researchers for further study.

## 4. Conclusions

Sn-substituted polycrystalline ferrites, NZSFO ($x$ = 0.0, 0.05, 0.1, 0.15, 0.2 and 0.3) sintered at 1300 °C, have been successfully synthesized using standard ceramic technique. The single phase spinel structure of the samples has been confirmed from the XRD patterns. The grain size increases from 7.8 to 34.8 µm with increasing Sn content. The saturation magnetization is found to be decreased with increasing Sn concentration while the coercivity is increased. The initial permeability is fairly constant up to 10 MHz i.e., wide range of operating frequency or stability region has been achieved for the samples. The Curie temperatures rise gradually with increasing Sn content, except for $x$ = 0.05, which is fruitfully explained by the variation of *A-B* interaction strength due to dopant cations entering in different sites. A reasonably low $H_c$ for $x$ = 0.0, implies that this material might be a promising candidate for transformer core and inductor applications.

**Acknowledgements:** The authors are grateful to the Directorate of Research and Extension, Chittagong University of Engineering and Technology (CUET), Chittagong-4349, Bangladesh for arranging the financial support for this work.